\RequirePackage{fix-cm}
\documentclass[smallextended]{svjour3}       
\smartqed  
\usepackage{bm}
\usepackage{graphicx}
\usepackage{amsfonts,amsmath,amssymb}
\usepackage[pdfstartview=FitH,colorlinks=true,linkcolor=blue,citecolor=blue,urlcolor=blue]{hyperref}
\begin{document}

\title{Superfluid Gap in Neutron Matter 
from a Microscopic Effective Interaction
}


\author{Omar Benhar         \and
        Giulia De Rosi 
}


\institute{Omar Benhar 
\at
INFN and Dipartimento di Fisica, Universit\`a ``La Sapienza'', Roma \\
              \email{omar.benhar@roma1.infn.it}           
           \and
           Giulia De Rosi 
           \at  INO-CNR BEC Center and Dipartimento di Fisica, Universit\`a di Trento \\
}

\date{Received: date / Accepted: date}

\maketitle

\begin{abstract}
Correlated Basis Function (CBF) perturbation theory and the
formalism of cluster expansions have been recently employed to
obtain an effective interaction from a nuclear Hamiltonian strongly
constrained by phenomenology.
We report the results of a study of the superfluid gap in pure neutron
matter, associated with the formation of Cooper pairs in the $^1S_0$ channel.
The calculations have been carried out
using an improved version of the CBF effective interaction, in which
three-nucleon forces are taken into account using a microscopic model.
Our results show that a non-vanishing superfluid gap develops
at densities in the range $2 \times 10^{-4}  \lesssim \rho/\rho_0 \lesssim 0.1 $,  where $\rho_0 = 2.8 \times 10^{14}$ g cm$^{-3}$ is the
equilibrium density of isospin-symmetric nuclear matter, 
corresponding mainly to the neutron star inner crust.
\keywords{Neutron stars \and Superfluidity \and Nuclear matter \and Correlated basis functions \and Cluster expansions}
\end{abstract}

\section{Introduction}
\label{Sec:Intro}
\indent

In neutron stars, both conditions for the occurrence of superfluidity in fermionic systems, 
that is, strong degeneracy and
the existence of an attractive interaction between the constituents of strongly interacting matter, are believed to be fulfilled \cite{Alford2006}.

The onset of a superfluid (and/or superconducting) phase does not have a significant impact on the
equation of state, determining  the equilibrium properties of the star, except in the
very low density region of the crust. The condensation energy\textemdash i.e. the difference between
the energies of the normal and superfluid  states associated with the formation of Cooper pairs \cite{Cooper1956}\textemdash
is in fact small, although not totally negligible, with respect to the typical energies of the normal phase \cite{Yang1971}.

The main effect of the superfluid transition is the opening of an energy gap 
at the Fermi surface \cite{March1967}. This leads to a reduction of the phase space available
to particles undergoing scattering processes, which in turn results in a strong modification of
the neutrino emission, scattering and absorption rates, as well as of the transport coefficients,
including the shear viscosity and thermal conductivity.
As a consequence, a quantitative understanding of the superfluid phase transition is required to study
both neutron-star cooling \cite{cooling} and the onset of the Chandrasekhar-Friedman-Schutz (CFS) instability of rotating stars \cite{Chandrasekhar1970,FS}, 
which is largely driven by dissipative processes \cite{Andersson2001,Andersson2005}.

The approach based on effective interactions has long been recognized as well suited for the development of a 
unified description of equilibrium and non-equilibrium properties of nuclear matter, 
based on realistic models of nuclear dynamics at microscopic level \cite{Yang1971,Amundsen1985,Wambach1993}.

In recent implementations, the effective interaction 
has been derived from realistic phenomenological Hamiltonian\textemdash strongly 
constrained by the available data\textemdash within the
formalism of Correlated Basis Functions (CBF) \cite{Cowell2006,Benhar2007,Lovato2011,Lovato2013,Lovato2014}. Unlike the bare nucleon-nucleon force, 
the effective
interaction is well behaved at short distances, and can be used to carry out perturbative
calculations in the basis of eigenstates of the non-interacting system.

Existing applications of the CBF effective interaction include calculations of the shear 
viscosity and thermal conductivity 
coefficients of neutron matter \cite{Benhar2007,Benhar2010}, as well as the nuclear matter response to neutrino interactions 
\cite{Cowell2006,Lovato2013,Lovato2014,Benhar2009,Benhar2013}.
The potential of the approach based on effective interactions obtained from correlated functions
has been recently confirmed by the results of systematic studies of the properties of the Fermi hard sphere system~\cite{Mecca2015,Mecca2016}, providing
a valuable model of nuclear matter.

In this work, we report the results of a calculation of the superfluid gap associated with the 
formation of Cooper pairs in the $^1S_0$ channel in pure neutron matter (PNM), performed using the CBF 
effective interaction derived in Refs.~\cite{Lovato2011,Lovato2013}.

The paper is organized as follows. In Sec. \ref{Sec:Formalism} we outline the {\em ab initio} approach based on a microscopic nuclear Hamiltonian, and discuss  
the derivation of an effective nucleon-nucleon (NN) interaction\textemdash suitable to carry out perturbative calculation using the basis states of the non-interacting system\textemdash performed combining the CBF formalism and the cluster expansion technique. 
The differences between the effective interaction and the bare NN potential are also illustrated. 
The numerical results, indicating the 
occurrence of a superfluid phase of PNM at densities corresponding to the inner crust of neutron stars,  are reported in Sec. \ref{Sec:Results}.
Finally, in Sec. \ref{Sec:Conclusions} we summarize our findings, and draw the perspectives for future application of the CBF effective interaction approach.

\section{Formalism}
\label{Sec:Formalism}
\indent
In this section, we briefly outline the phenomenological model of nuclear dynamics employed in our work, and describe the procedure leading to 
the determination of the effective interaction.

\subsection{The nuclear Hamiltonian}
\label{hamiltonian}
\indent

The formalism of nuclear many-body theory provides a consistent framework, suitable for 
treating the non-perturbative nature of NN interactions. 
Within this approach, nuclear matter is modelled as a collection of  point-like particles, the dynamics of which 
are dictated by the Hamiltonian
\begin{equation}
\label{Eq:H}
H=\sum_{i}\frac{{\bf p}_i^2}{2m }+\sum_{j>i}v_{ij}+\ldots \ ,
\end{equation}
where ${\bf p}_i$ and $m$ denote the momentum of the $i$-th nucleon and its mass, respectively, $v_{ij}$ is the 
NN interaction potential and the ellipses refer to the presence of {\em irreducible} interactions involving three or more nucleons.
The inclusion of a three-nucleon potential, $V_{ijk}$,  is in fact necessary to explain the properties of the three-nucleon systems, 
as well as saturation of isospin-symmetric nuclear matter (SNM). 

The NN potential $v_{ij}$
reduces to the Yukawa one-pion exchange potential at large
distances, while its behavior at short and intermediate range is determined by
a fit of deuteron properties and NN scattering phase shifts.

Coordinate-space NN potentials are usually written in the form
\begin{equation}
v_{ij}=\sum_{p} v^{p}(r_{ij}) O^{p}_{ij} \ , 
\label{eq:NN_1}
\end{equation}
where $r_{ij} = |{\bf r}_i - {\bf r}_j|$ is the distance between the interacting particles, and the sum includes up to eighteen terms. 
The most prominent contributions are  those associated with the operators 
\begin{align}
O^{p \leq 6}_{ij} = [1, (\boldsymbol{\sigma}_{i}\cdot\boldsymbol{\sigma}_{j}), S_{ij}]
\otimes[1,(\boldsymbol{\tau}_{i}\cdot\boldsymbol{\tau}_{j})]  \ ,
\label{av18:2}
\end{align}
where $\boldsymbol{\sigma}_{i}$ and $\boldsymbol{\tau}_{i}$ are Pauli matrices acting in spin and isospin space, respectively, while the operator 
\begin{align}
S_{ij}=\frac{3}{r_{ij}^2}
(\boldsymbol{\sigma}_{i}\cdot{\bf r}_{ij}) (\boldsymbol{\sigma}_{j}\cdot{\bf r}_{ij})
 - (\boldsymbol{\sigma}_{i}\cdot\boldsymbol{\sigma}_{j}) \ , 
 \label{S12}
\end{align}
reminiscent of the potential describing the interaction between two magnetic dipoles, accounts for the occurrence of non-spherically-symmetric forces. 

The potential models obtained including the six operators of Eqs.~\eqref{av18:2}-\eqref{S12} 
explain deuteron properties and the $S$-wave scattering phase shifts up to pion production threshold. 

\subsection{The CBF effective interaction}
\label{CBF:veff}
\indent

 Performing perturbative calculations in the basis of eigenstates of the non-interacting system requires the replacement of the  \emph{bare} NN potential\textemdash featuring a strongly repulsive core\textemdash with a well-behaved effective 
interaction \cite{Ring1980,Fetter2012}, that can be obtained either summing up ladder diagrams at all orders, as in $G$-matrix perturbation theory, \cite{Wambach1993,Benhar2010,Zhang2010} 
or modifying the basis states, as in the CBF approach~\cite{Cowell2006,Benhar2007,Benhar2010}.

Within the CBF formalism, non-perturbative effects are taken into account replacing the states of 
the non-interacting system, i.e. the Fermi gas states $| n_{\rm FG} \rangle$ in the case of 
 uniform nuclear matter,  with a set of {\em correlated} states, defined as (see, e.g., Refs. \cite{Feenberg2012,Pandharipande1979,Clark1979})
\begin{equation}
\label{Eq:corrbasis}
| n \rangle =\frac{F | n_{\rm FG} \rangle}{\langle n_{\rm FG}| F^{\dagger} F | n_{\rm FG} \rangle^{1/2}} \ .
\end{equation}
The operator $F$, embodying the correlation structure induced by the NN
interaction, is written in the form
\begin{equation}
\label{Eq:F}
F=\mathcal{S}\prod_{j>i} f_{ij} \  ,
\end{equation}
with
\begin{align}
f_{ij}=\sum_p f^p(r_{ij}) O^{p}_{ij}  \  , 
\label{def:fij}
\end{align}
the two-body operators $O^{p}_{ij}$ being the same as in Eq.\eqref{eq:NN_1}.
Note that, because the operator structure of $f_{ij}$ reflects the complexity of the NN potential, the product 
appearing in Eq.\eqref{Eq:F} needs to be  symmetrysed through the action of the operator $\mathcal{S}$, to account 
for the fact that $[f_{ij},f_{ik}] \neq 0$. 

In principle, the radial  dependence of the correlation functions $f^p(r_{ij})$ can be determined from
functional minimization of the expectation value of the Hamiltonian in the correlated ground state
\begin{equation}
\label{EV}
E_V = \langle 0 | H | 0 \rangle \ .
\end{equation}
In practice, however, the calculation of the variational energy of Eq.~\eqref{EV} involves non trivial  difficulties. It can be effectively  carried out
expanding the right-hand side in a series, whose terms describe the contributions of subsystems, or
clusters, involving an increasing number of correlated particles (see, e.g., Refs.~\cite{Clark1979,Fabrocini1986}). The terms of the cluster expansion can be represented by diagrams, that
are  classified according to their topological structures.  Selected classes of diagrams can then be summed up to all orders solving a set of coupled
non-linear integral equations\textemdash referred to as Fermi Hyper-Netted Chain/Single-Operator Chain (FHNC/SOC) equations~\cite{Pandharipande1979,Fantoni1974}\textemdash to obtain an accurate estimate of the ground state energy.
The full derivation of the Euler equation obtained from 
\begin{align}
\label{euler1}
\frac{\delta E_V}{\delta F} = 0 \ ,
\end{align}
within the FHNC scheme is discussed in Ref.\cite{euler}.

The new basis defined by Eq. \eqref{Eq:corrbasis} 
can be employed to perform perturbative calculations with the {\em bare} NN potential, although the non-orthogonality of the basis states entails
severe computational difficulties~\cite{Fantoni1988}. However, the same formalism can be also exploited to obtain 
an {\em effective interaction}, suitable to be used with the 
Fermi gas basis \cite{Cowell2006,Benhar2007}. 

The CBF effective interaction, $v^{\rm eff}$, is defined through the relation \cite{Cowell2004}
\begin{equation}
\label{Eq:def_veff}
\langle 0 | H | 0 \rangle =
\langle 0_{\rm FG} |  \ \sum_{i}\frac{{\bf p}_i^2}{2m }+\sum_{j>i}v^{\rm eff}_{ij} \ | 0_{\rm FG} \rangle \ ,
\end{equation}
where $| 0_{\rm FG} \rangle$ and $| 0 \rangle$ denote the Fermi gas and CBF ground state, respectively, 
and $H$ is the nuclear Hamiltonian of Eq. \eqref{Eq:H}. 

In the pioneering works of Refs.~\cite{Cowell2006,Benhar2007} the left hand side of Eq. \eqref{Eq:def_veff} has been evaluated 
using a truncated version of the state-of-the-art Argonne $v_{18}$ potential, including contributions with $p \leq 6$ [see 
Eqs.\eqref{eq:NN_1}-\eqref{S12}]\cite{Wiringa1995,Pudliner1997} and including two-nucleon cluster contributions only. 
This approximation leads to the simple expression 
\begin{align}
\label{Eq:def_veff2}
v^{\rm eff}_{ij}  = \frac{1}{m} \left( \bm{\nabla} f_{ij} \right)^2 +  f_{ij}   v_{ij} f_{ij} ,
\end{align}
where $v_{ij}$ is the bare NN potential and the $f_{ij}$ are determined solving the Euler equations derived from 
the approximated energy functional, with the correlation range fixed in such a way as to reproduce the 
FHNC/SOC results  obtained with the same Hamiltonian. In Ref.~\cite{Benhar2007}, the effects of three- and many-nucleon interactions 
have been also taken into account, using a density-dependent modification of the NN potential originally proposed in Ref.~\cite{Lagaris1981}.

More recently, an improved CBF effective interaction has been derived by the authors of Refs. \cite{Lovato2011,Lovato2013,Lovato2014}, who
explicitly included three-body cluster contributions to the left hand side of Eq.\eqref{Eq:def_veff}. This scheme allows for 
a more realistic treatment of three-body forces, which are known to play a critical role in determining
both the spectra of few-nucleon systems 
and the saturation properties of SNM, based on a realistic description at microscopic level.

The CBF effective interaction of Refs.~\cite{Lovato2011,Lovato2013} has been 
obtained from a nuclear Hamiltonian comprising the Argonne $v_6^\prime$ NN
potential \cite{Pudliner1997,Wiringa2002} and the Urbana IX (UIX) three-nucleon potential \cite{Pudliner1995}.
The $v_6^\prime$ potential accounts for deuteron properties and $S$-wave NN scattering phase shifts, while
the UIX potential, including a Fujita-Miyazawa two-pion exchange attractive term \cite{Fujita1957} and a purely phenomenological repulsive term, 
is designed to reproduce the properties of the three-nucleon bound states and the saturation density of SNM.

\begin{figure*}
\centering
\includegraphics[width=0.80\textwidth]{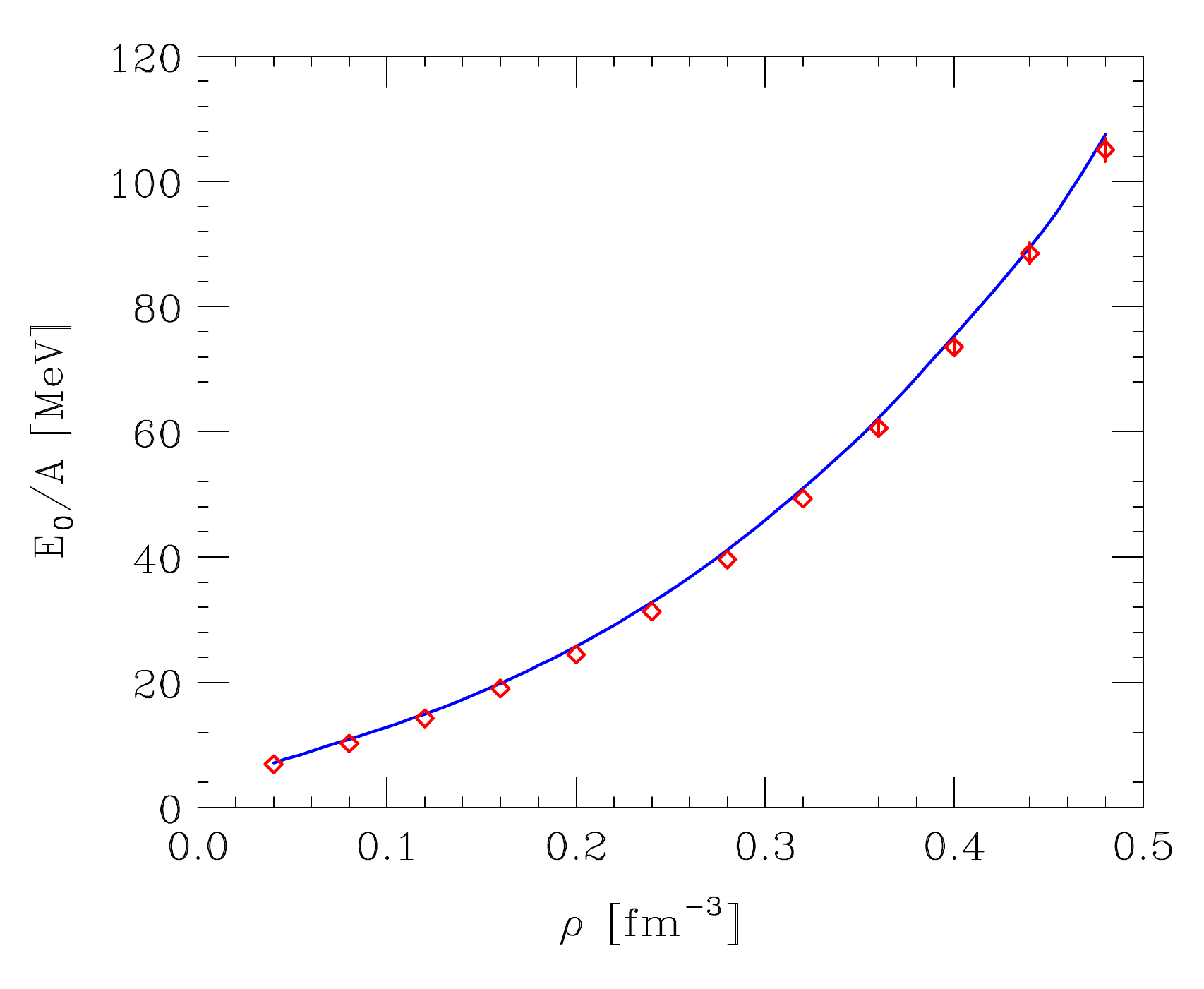}[h]
\caption{Energy per particle of PNM as a function of the density, $\rho$. The solid line represents the energies obtained using the CBF effective interaction\textemdash which coincide with 
the FHNC/SOC energies by construction\textemdash while the circles correspond to the values calculated using the Auxiliary Field Diffusion Monte Carlo (AFDMC) technique \cite{Schmidt1999}. Note that the statistical error bars associated with the Monte Carlo energies are only visible $\rho \gtrsim 0.4$ {\rm fm}$^{-3}$.}
\label{fig:EoS_PNM} 
\end{figure*}

The energy per particle of both isospin-symmetric nuclear SNM and PNM 
obtained from $v^{\rm eff}$ in the Hartree-Fock approximation\textemdash which reproduces by construction the the FHNC/SOC variational results computed using the 
$v_6^\prime$+UIX Hamiltonian\textemdash 
turns out to also be in excellent  agreement with the results of other highly advanced many-body approaches \cite{Lovato2013}. 
Figure~\ref{fig:EoS_PNM}, showing a comparison with the PNM energies obtained using the Auxiliary Field Diffusion Monte Carlo technique \cite{Schmidt1999},
strongly suggests that the FHNC/SOC scheme provides a very accurate upper bound to the ground state energy.

\begin{figure*}
\centering
\includegraphics[width=0.75\textwidth]{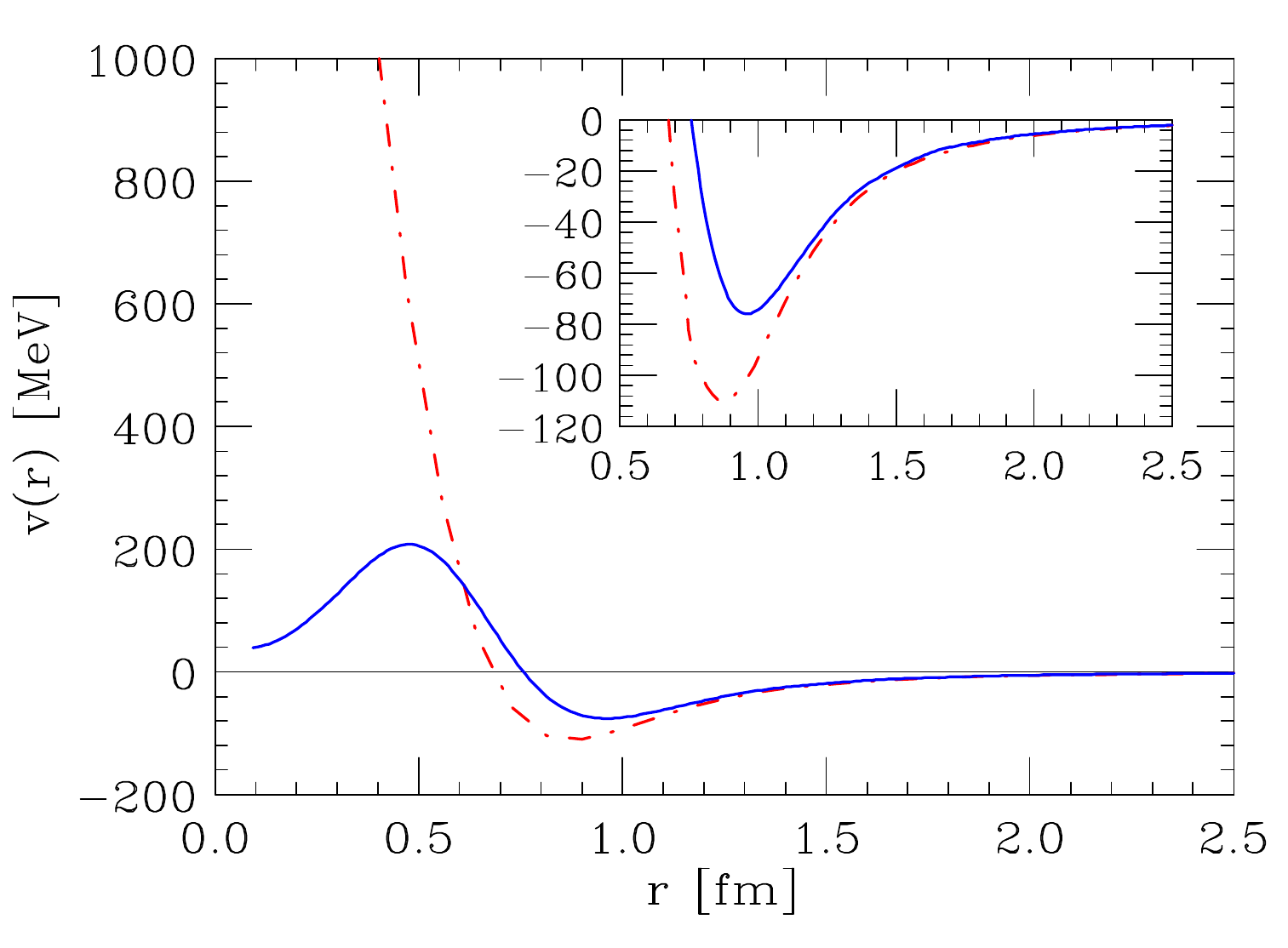}[h]
\caption{Nucleon-nucleon potential in the $S =0$, $T=1$ channel. The solid and dot-dash lines
correspond to the CBF effective interaction of Ref. \cite{Lovato2013} and to the bare
Argonne $v_6^\prime$ \cite{Pudliner1997} potential, respectively. The inset shows a blow up of the region 
$0.5 \leq r \leq 2.5$ fm}.
\label{fig:potential} 
\end{figure*}

The solid line of Fig. \ref{fig:potential} shows the radial dependence of the CBF effective interaction 
in the $S =0$, $T=1$ channel, $S$ and $T$ being the total spin and isospin of the interacting pair, 
respectively. 
Comparison with the dot-dash line, corresponding to the bare $v_6^\prime$ 
potential, clearly illustrates the screening effect arising from NN correlations, 
leading to the near-disappearance of the short-range repulsive core of the bare interaction.
In addition, due to the modification of the two-nucleon wave function arising from the inclusion of correlations,  the effective interaction
includes an additional purely kinetic term. The inset shows a blow up of the attractive region. 

\section{Results}
\label{Sec:Results}
\indent

The generalization of the formalism originally derived by Bardeen, Cooper and Schrieffer~\cite{Bardeen1957} to 
 allow the use of correlated basis functions is thoroughly discussed in Ref.~\cite{Krotscheck1980}. 
For any baryon density $\rho = k_{\rm F}^3/(3\pi^2)$, where $k_{\rm F}$ denotes the Fermi momentum, the 
gap equation corresponding to $S$-wave coupling in cold PNM 
\begin{equation}
\label{Eq:gap}
\Delta(k) = -\frac{1}{\pi} \ \int {k^\prime}^2 dk^\prime \frac{ v(k,k^\prime) \Delta(k^\prime) }
{ \left[ \xi^2(k^\prime) + \Delta^2(k^\prime) \right]^{1/2} } \ ,
\end{equation}
has been solved using the algorithm discussed in Ref. \cite{Khodel1996}.
Eq. \eqref{Eq:gap} involves the momentum-space matrix elements of the potential
\begin{equation}
\label{Eq:v}
v(k,k^\prime) = \int r^2 dr j_0(kr) v^{\rm eff}(r) j_0(k^\prime r) \ , 
\end{equation}
where $j_0(x) = \sin(x)/x$ is the $0$-th order spherical Bessel function, $v^{\rm eff}(r)$ is the projection of the CBF effective potential 
in the $S =0$, $T=1$ channel (see Fig. \ref{fig:potential}), 
and
\begin{equation}
\label{Eq:xi}
\xi(k) = e(k) - \mu \ ,
\end{equation}
where $e(k)$ and $\mu=e(k_{\rm F})$ denote the energy of a particle carrying momentum $k$ and 
the chemical potential, respectively.

\begin{figure*}
\centering
\includegraphics[width=0.75\textwidth]{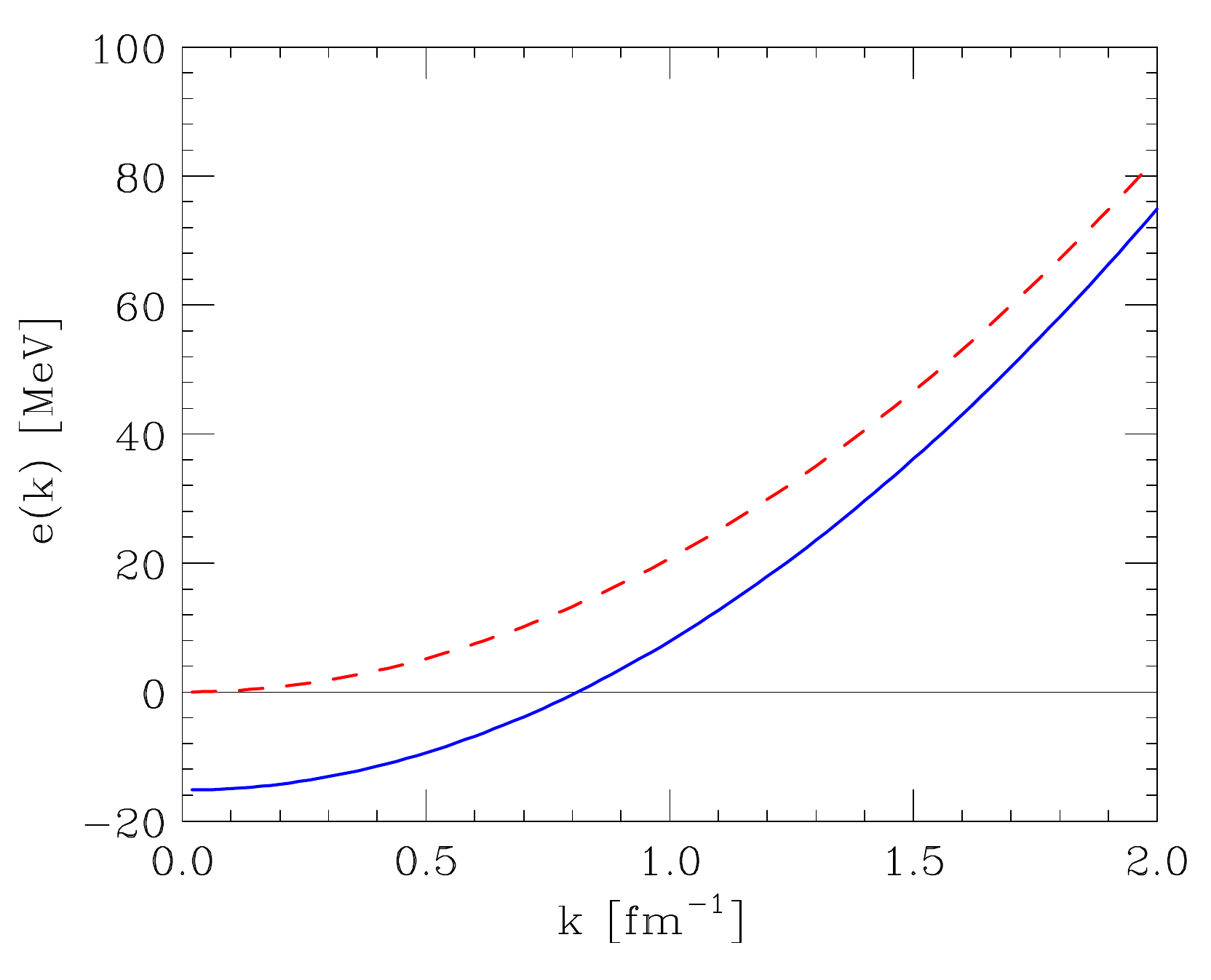}[h]
\caption{Momentum dependence of the single-particle spectrum in PNM. Solid line: results obtained at first order in the 
CBF effective interaction, corresponding to the Hartree-Fock approximation, at $\rho = 0.04$ fm$^{-1}$. Dashed line: kinetic energy 
spectrum.}
\label{fig:spectrum} 
\end{figure*}

The calculation has been carried out using the CBF effective interaction of Refs. \cite{Lovato2011,Lovato2013}.
The single-particle spectrum $e(k)$ has been consistently computed at first order in the CBF 
effective interaction, that is,  within the Hartree-Fock approximation. The solid line of Fig. \ref{fig:spectrum} illustrates the momentum
dependence of the Hartree-Fock spectrum of PNM at density $\rho = 0.04$ fm$^{-1}$. 
For comparison,  the kinetic energy spectrum is also shown, by the dashed line.

\begin{figure*}
\centering
\includegraphics[width=0.75\textwidth]{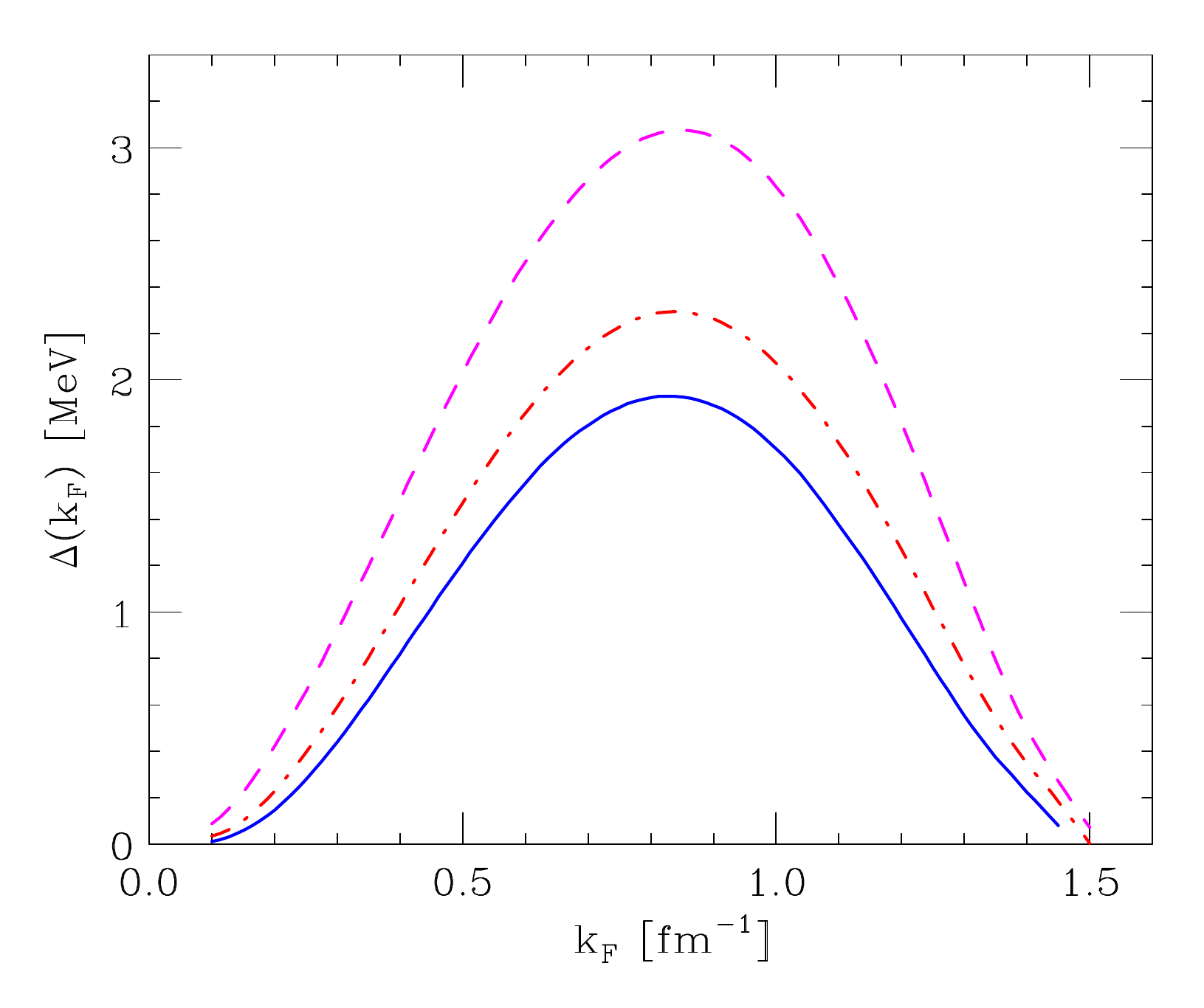}[h]
\caption{Fermi momentum dependence of the superfluid gap at the Fermi surface, $\Delta(k_{\rm F})$. 
The dashed line represents the results obtained using the bare $v_6^\prime$ potential and the kinetic energy spectrum, while the solid line corresponds to
calculations carried out using the CBF effective interaction and the Hartree-Fock spectrum. 
For comparison, the dash-dot line shows $\Delta(k_{\rm F})$ computed combining the CBF effective interaction with the kinetic energy spectrum.}
\label{fig:gap} 
\end{figure*}

The main results of our work are summarized in Fig. \ref{fig:gap}, showing the superfluid gap at the Fermi surface, $\Delta(k_{\rm F})$,   
as a function of the Fermi momentum $k_{\rm F}$. The solid line has been obtained using the CBF effective interaction and the
Hartee-Fock spectrum, while the dashed line corresponds to a calculation carried out with the bare $v_6^\prime$ NN
potential and the kinetic energy spectrum. The comparison shows that, while the range of Fermi momentum in which $\Delta(k_{\rm F}) \neq 0$ 
is about the same, the inclusion of interaction and correlation effects leads to a significant reduction of the gap. 

To clarify the roles played by the interaction employed to evaluate the matrix element entering Eq.\eqref{Eq:gap} 
and of the single-particle energies appearing in \eqref{Eq:xi}, the 
results obtained combining the CBF effective interaction with the kinetic energy spectrum are also displayed, by the dot-dash line.
It appears that while being sizable, the effect of interactions in the single particle spectrum is not as large as that arising from the replacement of 
the bare potential with the CBF effective interaction in Eq.\eqref{Eq:v}. In this context, it has to be also kept in mind that second order contributions, 
leading to the appearance of an explicit energy dependence of the neutron self-energy, are known to significantly affect $e(k)$ and the 
nucleon effective mass in the vicinity of the Fermi surface. The impact of these corrections on the determination of $\Delta(k)$ within the proposed approach 
should be carefully investigated. 

\begin{figure*}
\centering
\includegraphics[width=0.75\textwidth]{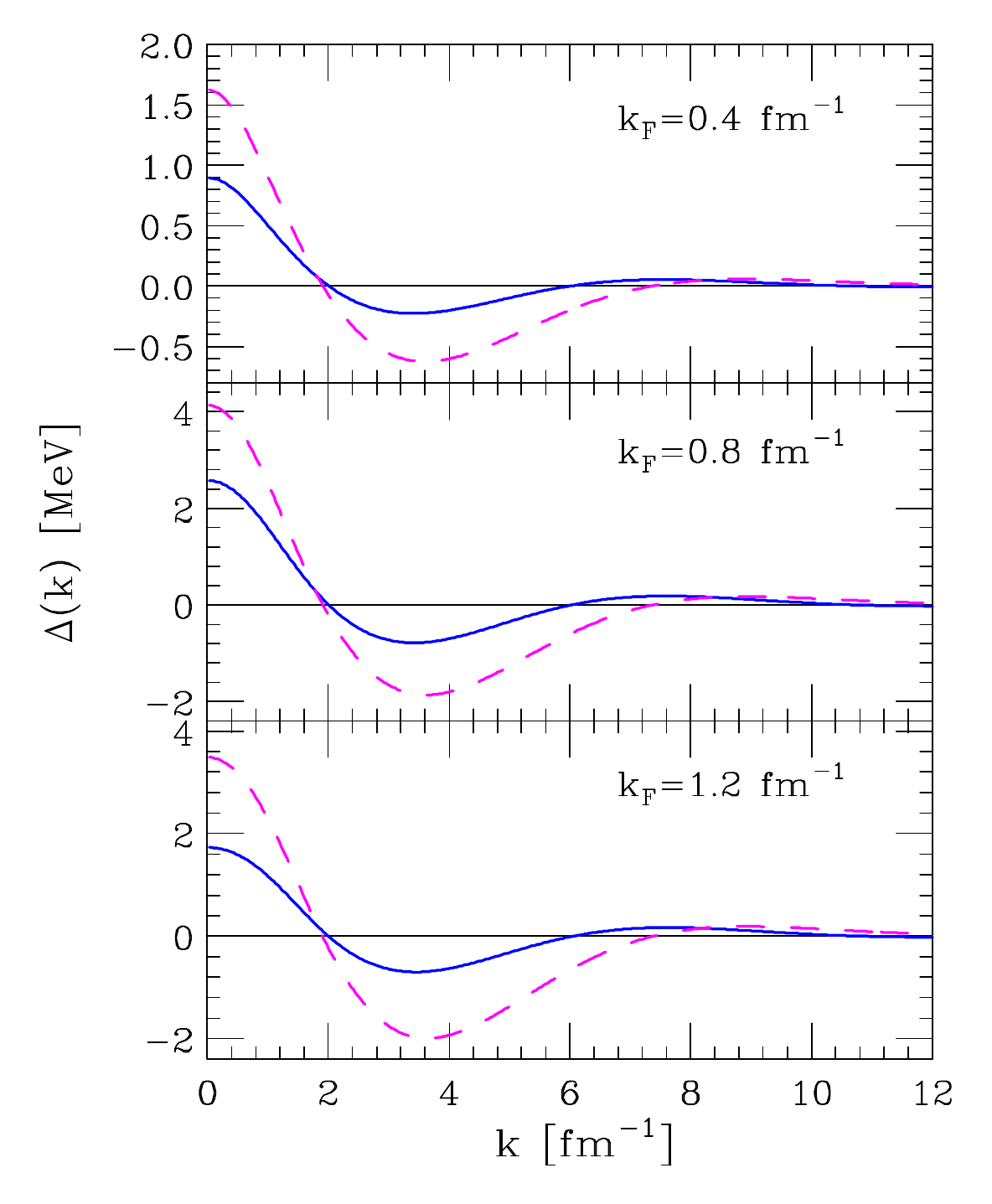}[h]
\caption{Momentum dependence of the superfluid gap of Eq.\eqref{Eq:gap}. 
The dashed line shows the
results obtained using the bare $v_6^\prime$ potential and the kinetic energy spectrum, while the solid line corresponds to
calculations carried out using the CBF effective interaction.}\label{fig:Delta_k} 
\end{figure*}

The solid lines of Fig.~\ref{fig:Delta_k} illustrate the momentum dependence of  the gap function, $\Delta(k)$, obtained using the CBF effective interaction 
for three different values of the Fermi momentum: $k_F=$ 0.4, 0.8 and 1.2 fm$^{-1}$. For comparison, the corresponding results obtained using the bare NN
potential and the kinetic energy spectrum are also shown, by the dashed lines.

As a final remark, we note that the present version of the code employed to obtain the numerical results shown in Fig.~\ref{fig:gap}, does not allow to pin down the contribution arising from 
three-nucleon interactions, unless three-nucleon cluster terms are disregarded altogether. However, in the low-density region in which 
the superfluid gap is nonzero, two nucleon interactions are expected to largely dominate.


\section{Summary and outlook}
\label{Sec:Conclusions}
\indent

We have carried out a calculation of the superfluid gap in PNM, associated with the formation
of Cooper pairs of neutrons in states of total spin $S=0$ and relative angular momentum $\ell=0$. The interaction
in this channel, which dominates the attractive component of the neutron-neutron force, has been described  within the CBF formalism, 
using an effective potential derived from the state-of-the-art phenomenological models of the two- and three-nucleon potentials
referred to as Argonne $v_6^\prime$ and UIX.

Note that the CBF effective interaction is {\em not} defined in operator form, but only in terms of
its expectation value in the Fermi gas ground state. However, unlike the Skyrme-like interactions 
derived using a similar procedure \cite{skyrme1,skyrme2}, it is strongly constrained by a microscopic model of nuclear dynamics.
Therefore, it is well suited to perform calculations of many different quantities, including the nucleon-nucleon  
scattering rates in matter, needed for a {\em consistent} description of equilibrium and non-equilibrium properties of 
neutron stars.

It is important to keep in mind that the validity of the assumption that perturbative calculations involving matrix elements of $v_{ij}^{\rm eff}$ between Fermi 
gas states provide accurate estimates of quantities other than the ground-state energy can not be taken for granted, and must be 
ultimately assessed at numerical level. A step along this line is the work of Refs.~\cite{skyrme1,skyrme2}, whose authors employed a CBF 
effective interaction to carry out calculations of a variety of properties of the Fermi hard-sphere system, ranging from 
the self-energy to the in-medium scattering cross section and the transport coefficients. The agreement between the results of this study and the 
predictions of low-density expansions appears to be quite encouraging.
 
We find that a non-vanishing superfluid gap develops in the density
range typical of the neutron star inner crust, extending from the the neutron drip density 
$\rho_{ND} \approx 4 \times 10^{11}$ g cm$^{-3}$ to $\rho \approx 10^{14}$ g cm$^{-3}$ \cite{crust}.

In the case of $^1S_0$ pairing, the critical temperature $T_c$ of the superfluid transition can be estimated from the
value of the gap at zero temperature ($T=0$) \cite{Tilley1990,Chen1993}. The resulting maximum value is in the range $T_c \sim 1-  2 \ {\rm MeV }$,
corresponding to $\sim 1 -  2 \ \times 10^{10} \ {\rm K}$.

Our results, while being interesting in their own right, should be regarded as a first step towards a 
comprehensive description of the superfluid and
superconductive phases of neutron stars. The interaction between neutrons coupled to total spin $S=1$ and angular momentum
$\ell=1$ is also attractive. The formation of Cooper pairs of neutrons with these quantum numbers is expected to occur at densities
$\rho > \rho_0$ typical of the neutron-star core. The appea\-rance of a superfluid phase in this region would strongly affect
the dissipative processes determining 
the stability of rotating stars. In addition, the small fraction -- typically less that $\sim 10 \%$ --
of protons are also expected to become superconductive, thus affecting the dissipative processes driven by 
electromagnetic interactions with electrons and muons.
The extension of the formalism employed in our work to study neutron superfluidity in the $^3P_2$ channel and proton superconductivity
does not involve any conceptual difficulties.


%
%

\begin{acknowledgements}
This research was supported by the Italian National Institute
for Nuclear Research (INFN) under grant MANYBODY.
\end{acknowledgements}


\begin{thebibliography}{}
%
%

\bibitem{Alford2006}
M. G. Alford, J. W. Clark, and A. Sedrakian, 
\textit{"Pairing in Fermionic Systems: Basic Concepts and Modern Applications"}.
World Scientific, Singapore (2006). 

\bibitem{Cooper1956}
L. N. Cooper, 
\textit{"Bound Electron Pairs in a Degenerate Fermi Gas"},
\href{https://journals.aps.org/pr/abstract/10.1103/PhysRev.104.1189}{Phys. Rev. \textbf{104}, 1189 (1956)}.


\bibitem{Yang1971}
C. - H. Yang and J. W. Clark, 
\textit{"Superfluid condensation energy of neutron matter"}, \href{http://www.sciencedirect.com/science/article/pii/0375947471910025}{Nucl. Phys. A \textbf{174}, 49 (1971)}.

\bibitem{March1967}
N. H. March, W. H. Young, and S. Sampanthar, 
\textit{"The Many-body Problem in Quantum Mechanics"}.
Dover Publications, New York (1967).


\bibitem{cooling}
D.G. Yakovlev and C.J Pethick, 
\textit{"Neutron Star Cooling"}
\href{http://www.annualreviews.org/doi/abs/10.1146/annurev.astro.42.053102.134013}{Ann. Rev. Astron. Astrophys. \textbf{42}, 169 (2004)}.

\bibitem{Chandrasekhar1970}
S. Chandrasekhar, 
\textit{"Solutions of Two Problems in the Theory of Gravitational Radiation"},
\href{https://journals.aps.org/prl/abstract/10.1103/PhysRevLett.24.611}{Phys. Rev. Lett. \textbf{24}, 611 (1970)}.

\bibitem{FS}
J.L. Friedman and B.F. Schutz, 
\textit{"Lagrangian perturbation theory of nonrelativistic fluids"}
\href{http://adsabs.harvard.edu/abs/1978ApJ...221L..99F}
{Astrophys. J. \textbf{221}, 937 (1978)}.

\bibitem{Andersson2001}
N. Andersson and K. D. Kokkotas, 
\textit{"The r-mode instability in rotating neutron stars"},
\href{http://www.worldscientific.com/doi/abs/10.1142/S0218271801001062}{Int. J. Mod. Phys. D \textbf{10}, 381 (2001)}.

\bibitem{Andersson2005}
N. Andersson, G.L. Comer, and K. Glampedakis, 
\textit{"How viscous is a superfluid neutron star core?"},
\href{http://www.sciencedirect.com/science/article/pii/S0375947405010572}{Nucl. Phys. A  \textbf{763}, 212 (2005)}.




\bibitem{Amundsen1985}
L. Amundsen and E. \O stgaard, 
\textit{"Superfluidity of neutron matter: (I). Singlet pairing"},
\href{http://www.sciencedirect.com/science/article/pii/S0375947485901034}{Nucl. Phys. A \textbf{437}, 487 (1985)}.


\bibitem{Wambach1993}
J. Wambach, T. L. Ainsworth, and D. Pines, 
\textit{"Quasiparticle interactions in neutron matter for applications in neutron stars"},
\href{http://www.sciencedirect.com/science/article/pii/037594749390317Q}{Nucl. Phys. A \textbf{555}, 128 (1993)}.



\bibitem{Cowell2006}
S. Cowell and V. R. Pandharipande,
\textit{"Weak interactions in hot nucleon matter"},
\href{https://journals.aps.org/prc/abstract/10.1103/PhysRevC.73.025801}{Phys. Rev. C \textbf{73}, 025801 (2006)}.


\bibitem{Benhar2007}
O. Benhar and M. Valli, 
\textit{"Shear Viscosity of neutron matter from realistic nucleon-nucleon interactions"},
\href{https://journals.aps.org/prl/abstract/10.1103/PhysRevLett.99.232501}{Phys. Rev. Lett. \textbf{99}, 232501 (2007)}.

\bibitem{Lovato2011}
A. Lovato, O. Benhar, S. Fantoni, A. Yu. Illarionov, and K. E. Schmidt, 
\textit{"Density-dependent nucleon-nucleon interaction from three-nucleon forces"},
\href{https://journals.aps.org/prc/abstract/10.1103/PhysRevC.83.054003}{Phys. Rev. C \textbf{83}, 054003 (2011)}.



\bibitem{Lovato2013}
A. Lovato, C. Losa, and O. Benhar, 
\textit{"Weak response of cold symmetric nuclear matter at three-body cluster level"},
\href{http://www.sciencedirect.com/science/article/pii/S0375947413000419#}{Nucl. Phys. A \textbf{901}, 22 (2013)}.


\bibitem{Lovato2014}
A. Lovato, O. Benhar, S. Gandolfi, and C. Losa,
\textit{"Neutral-current interactions of low-energy neutrinos in dense neutron matter"},
\href{https://journals.aps.org/prc/abstract/10.1103/PhysRevC.89.025804}{Phys. Rev. C \textbf{89}, 025804 (2014)}.

\bibitem{Benhar2010}
O. Benhar, A. Polls, M. Valli, and I. Vida\~na, 
\textit{"Microscopic calculations of transport properties of neutron matter"},
\href{https://journals.aps.org/prc/abstract/10.1103/PhysRevC.81.024305}{Phys. Rev. C \textbf{81}, 024305 (2010)}.

\bibitem{Benhar2009}
O. Benhar and N. Farina, 
\textit{"Correlation effects on the weak response of nuclear matter"},
\href{http://www.sciencedirect.com/science/article/pii/S0370269309010661}{Phys. Lett. B \textbf{680}, 305 (2009)}.


\bibitem{Benhar2013}
O. Benhar, A. Cipollone, and A. Loreti, 
\textit{"Weak response of neutron matter at low momentum transfer
"},
\href{https://journals.aps.org/prc/abstract/10.1103/PhysRevC.87.014601}{Phys. Rev C \textbf{87}, 014601 (2013)}.


\bibitem{Mecca2015}
A. Mecca, A. Lovato, O. Benhar, and A. Polls, 
\textit{"Effective-interaction approach to the Fermi hard-sphere system"},
\href{https://journals.aps.org/prc/abstract/10.1103/PhysRevC.91.034325}{Phys. Rev. C \textbf{91}, 034325 (2015)}.

\bibitem{Mecca2016}
A. Mecca, A. Lovato, O. Benhar, and A. Polls, 
\textit{"Transport properties of the Fermi hard-sphere system"},
\href{https://journals.aps.org/prc/abstract/10.1103/PhysRevC.93.035802}{Phys. Rev. C \textbf{93}, 035802 (2016)}.

\bibitem{Ring1980}
P. Ring and P. Schuck,
\textit{"The Nuclear Many-Body Problem"}.
Springer-Verlag, Berlin Heidelberg (1980).

\bibitem{Fetter2012}
A. L. Fetter and J. D. Walecka, 
\textit{"Quantum Theory of Many-Particle Systems"}.
Dover Publications, New York (2012).

\bibitem{Zhang2010}
H. F. Zhang, U. Lombardo, and W. Zuo,
\textit{"Transport parameters in neutron stars from in-medium 
$NN$ cross sections"},
\href{https://journals.aps.org/prc/abstract/10.1103/PhysRevC.82.015805}{Phys. Rev. C \textbf{82}, 015805 (2010)}.

\bibitem{Feenberg2012}
E. Feenberg, 
\textit{"Theory of Quantum Fluids"}.
Academic Press, New York (1969).

\bibitem{Pandharipande1979}
V. R. Pandharipande and R. B. Wiringa,
\textit{"Variations on a theme of nuclear matter"},
\href{https://journals.aps.org/rmp/abstract/10.1103/RevModPhys.51.821}{Rev. Mod. Phys. \textbf{51}, 821 (1979)}.

\bibitem{Clark1979}
J. W. Clark, 
\textit{"Variational theory of nuclear matter"},
\href{http://www.sciencedirect.com/science/article/pii/0146641079900048}{Prog. Part. Nucl. Phys. \textbf{2}, 89 (1979)}.

\bibitem{Fabrocini1986}
A. Fabrocini and S. Fantoni, 
\textit{"First International Course of Condensed Matter
Physics"}, 
eds. D. Prosperi, S. Rosati and G. Violini.
World Scientific, Singapore (1986).

\bibitem{Fantoni1974}
S. Fantoni and S. Rosati, 
\textit{"Jastrow correlations and an irreducible cluster expansion for infinite boson or fermion systems"},
\href{https://link.springer.com/article/10.1007/BF02727446}{Il Nuovo Cimento A \textbf{20}, 179 (1974)}.

\bibitem{euler}
E. Krotscheck, 
\textit{"Fermi-Hypernetted Chain Theory for Liquid 3He: A Reassessment"},
\href{https://link.springer.com/article/10.1023/A\%3A1004664619961}{J.~Low Temp.  Phys. \textbf{119}, 103 (2000)}.

\bibitem{Fantoni1988}
S. Fantoni and V. R. Pandharipande, 
\textit{"Orthogonalization of correlated states"},
\href{https://journals.aps.org/prc/abstract/10.1103/PhysRevC.37.1697}{Phys. Rev. C \textbf{37}, 1697 (1988)}.

\bibitem{Cowell2004}
S. Cowell and V. R. Pandharipande,
\textit{"Neutrino mean free paths in cold symmetric nuclear matter"},
\href{https://journals.aps.org/prc/abstract/10.1103/PhysRevC.70.035801}{Phys. Rev. C \textbf{70}, 035801 (2004)}.

\bibitem{Wiringa1995}
R. B. Wiringa, V. G. J. Stoks, and R. Schiavilla,
\textit{"Accurate nucleon-nucleon potential with charge-independence breaking"},
\href{https://journals.aps.org/prc/abstract/10.1103/PhysRevC.51.38}{Phys. Rev C \textbf{51}, 38 (1995)}.

\bibitem{Pudliner1997}
B. S. Pudliner, V. R. Pandharipande, J. Carlson, S. C. Pieper, and R. B. Wiringa, 
\textit{"Quantum Monte Carlo calculations of nuclei with $A <\sim 7$"},
\href{https://journals.aps.org/prc/abstract/10.1103/PhysRevC.56.1720}{Phys. Rev. C \textbf{56}, 1720 (1997)}.

\bibitem{Lagaris1981}
I. E. Lagaris and V. R. Pandharipande, 
\textit{"Variational calculations of realistic models of nuclear matter"},
\href{http://www.sciencedirect.com/science/article/pii/0375947481902414}{Nucl. Phys. A \textbf{359}, 349 (1981)}.

\bibitem{Wiringa2002}
R. B. Wiringa and S. C. Pieper,
\textit{"Evolution of Nuclear Spectra with Nuclear Forces"},
\href{https://journals.aps.org/prl/abstract/10.1103/PhysRevLett.89.182501}{Phys. Rev. Lett. \textbf{89}, 182501 (2002)}.

\bibitem{Pudliner1995}
B. S. Pudliner, V. R. Pandharipande, J. Carlson, and R. B. Wiringa,
\textit{"Quantum Monte Carlo Calculations of $A \leq 6$ Nuclei"},
\href{https://journals.aps.org/prl/abstract/10.1103/PhysRevLett.74.4396}{Phys. Rev. Lett. \textbf{74}, 4396 (1995)}.

\bibitem{Fujita1957}
J.-i. Fujita and H. Miyazawa,
\textit{"Pion Theory of Three-Body Forces"},
\href{https://academic.oup.com/ptp/article-lookup/doi/10.1143/PTP.17.360}{Prog. Theor. Phys. \textbf{17}, 360 (1957)}.

\bibitem{Schmidt1999}
K. E. Schmidt and S. Fantoni, 
\textit{"A quantum Monte Carlo method for nucleon systems"},
\href{http://www.sciencedirect.com/science/article/pii/S0370269398015226}{Phys. Lett. B \textbf{446}, 99 (1999)}.

\bibitem{Bardeen1957}
J. Bardeen, L. N. Cooper, and J. R. Schrieffer,
\textit{"Theory of Superconductivity"},
\href{https://journals.aps.org/pr/abstract/10.1103/PhysRev.108.1175}{Phys. Rev. \textbf{108}, 1175 (1957)}.

\bibitem{Krotscheck1980}
E. Krotscheck and J. W. Clark,  
\textit{"Studies in the method of correlated basis functions: (III). Pair condensation in strongly interacting Fermi systems"},
\href{http://www.sciencedirect.com/science/article/pii/0375947480900172}{Nucl. Phys. A  \textbf{333}, 77 (1980)}.


\bibitem{Khodel1996}
V. A. Khodel, V. V. Khodel, and J. W. Clark, 
\textit{"Solution of the gap equation in neutron matter"},
\href{http://www.sciencedirect.com/science/article/pii/0375947495004777}{Nucl. Phys. A \textbf{598}, 390 (1996)}.

\bibitem{Tilley1990}
D.R. Tilley and J. Tilley, 
\textit{"Superfluidity and Superconductivity"}.
CRC Press, Boca Raton (1990).

\bibitem{Chen1993}
J. M. C. Chen, J. W. Clark, R. D. Dav\'e, and V. V. Khodel, 
\textit{"Pairing gaps in nucleonic superfluids"},
\href{http://www.sciencedirect.com/science/article/pii/037594749390314N}{Nucl. Phys. A  \textbf{555}, 59 (1993)}.

\bibitem{skyrme1}
J. Rikovska Stone, J. C. Miller, R. Koncewicz, P. D. Stevenson, and M. R. Strayer,
\textit{"Nuclear matter and neutron-star properties calculated with the Skyrme interaction"},
\href{https://journals.aps.org/prc/abstract/10.1103/PhysRevC.68.034324}{Phys. Rev. C \textbf{68}, 034324 (2003)}.

\bibitem{skyrme2}
E.Chabanat, P.Bonche, P. Haensel, J. Meyer, and R. Schaeffer, 
\textit{"A Skyrme parametrization from subnuclear to neutron star densities"},
\href{http://www.sciencedirect.com/science/article/pii/S0375947497005964}{Nucl. Phys. A \textbf{627}, 710 (1997)}.

\bibitem{crust}
N. Chamel and P. Haensel, 
\textit{"Physics of Neutron Star Crusts"},
\href{https://link.springer.com/article/10.12942/lrr-2008-10}{Living Reviews in Relativity \textbf{11}, 10 (2008)}.



 
 

























\end{thebibliography}


\end{document}